\documentclass[sigconf]{acmart}

\AtBeginDocument{%
  \providecommand\BibTeX{{%
    \normalfont B\kern-0.5em{\scshape i\kern-0.25em b}\kern-0.8em\TeX}}}

\copyrightyear{2020}
\acmYear{2020}
\setcopyright{none}
\acmConference[Author Preprint]{Author Preprint}
\acmBooktitle{Arxiv}
\acmDOI{.../...}
\acmISBN{-}

\usepackage{listings}
\lstdefinestyle{mystyle}{
    basicstyle=\ttfamily\small,
    numbers=left,
}
\begin{document}

\title{Automated System Performance Testing at MongoDB}

\author{Henrik Ingo}
\email{henrik.ingo@mongodb.com}
\orcid{0000-0003-1571-5108}
\affiliation{%
  \institution{MongoDB Inc}
}

\author{David Daly}
\email{david.daly@mongodb.com}
\orcid{0000-0001-9678-3721}
\affiliation{%
  \institution{MongoDB Inc}
}


\begin{abstract}
Distributed Systems Infrastructure (DSI) is MongoDB's framework for running
fully automated system performance tests in our Continuous Integration (CI)
environment. To run in CI it needs to automate everything end-to-end:
provisioning and deploying multi-node clusters, executing tests, tuning the
system for repeatable results, and collecting and analyzing the results.
Today DSI is MongoDB's most used and most useful performance testing tool.
It runs almost 200 different benchmarks in daily CI, and we also use it for
manual performance investigations. As we can alert the responsible engineer
in a timely fashion, all but one of the major regressions were fixed before
the 4.2.0 release. We are also able to catch net new
improvements, of which DSI caught 17. We open sourced DSI in March 2020.
\end{abstract}



\begin{teaserfigure}
  \includegraphics[width=\textwidth]{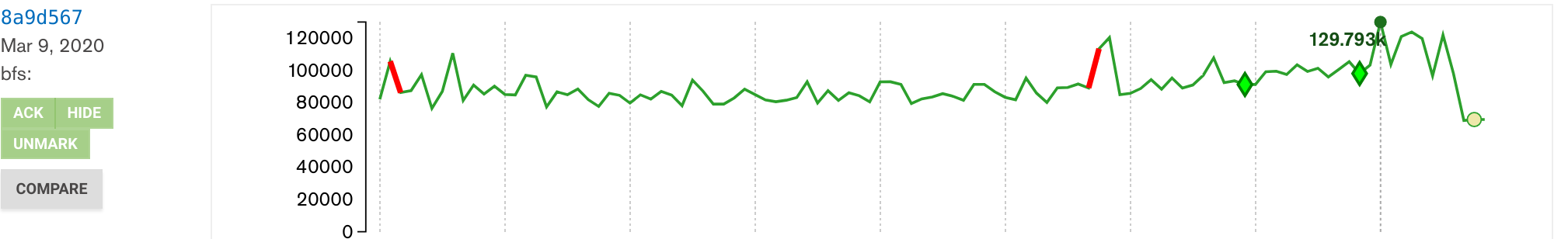}
  \caption{Timeseries of daily build results for YCSB load}
  \Description{Timeseries of daily build results for YCSB load}
  \label{fig:teaser}
\end{teaserfigure}

\maketitle

\section{Introduction}
MongoDB created a dedicated performance testing team in 2013. The first
approach was to build a team of systems engineers and performance experts
that would manually benchmark new releases of MongoDB, and was found
to be problematic. It was not scalable, and it also had issues with
repeatability.

The focus then shifted to running system performance tests in a daily Continuous Integration (CI)
build, just like is done for any other testing. But where all our other
tests can run on a single server, the system performance builds needed
to deploy realistic MongoDB clusters in EC2.
Initially this was accomplished by moving some shell
scripts, and Terraform\cite{Terraform} config files, from
the manual benchmarking, into the CI configuration. 

From these humble beginnings arose a software development project to
design and rewrite the same system more properly in Python. The framework
was called \textit{Distributed System Infrastructure}, or \textbf{DSI}
for short.

In this paper, we present the architecture and design of this framework
and results from using it to benchmark MongoDB. DSI is used in MongoDB
to run hundreds of automated benchmarks per day. The benchmarks in CI
are fully automated end-to-end: provisioning EC2 infrastructure and
deploying the MongoDB cluster, orchestrating test execution, and
collecting and analyzing results. It is also used by engineers to create
new tests, reproduce results from CI and manual "explorative"
benchmarking. Since 2018, DSI is our most used and most useful performance
testing tool. We have open sourced it in March 2020~\cite{DSIRepo}.

\section{Problem Definition}
The focus for DSI was serving the more complex requirements of
\textit{end-to-end system performance tests on real clusters, automating
every step including provisioning of hardware, and generating consistent,
repeatable results}.

More specifically, the high level goals of the project were:

\begin{enumerate}
    \item Full end-to-end automation.
    \item Support both CI and manual benchmarking.
    \item Elastic, public cloud infrastructure.
    \item Everything configurable.
    \item All configuration is via a single, consistent, YAML based system.
    \item Clear modularity for executables and configuration.
    \item Diagnosability.
    \item Repeatability.
\end{enumerate}

\subsection{Configuration} \label{sec:problem-configuration}
The main design goal was to move from a model where each new configuration
required a new shell script, to a model where everything was
driven by configuration files and no code changes would be needed to add a
new MongoDB configuration or test.

Further, even for parts that had configuration files, it was becoming
tedious to deal with so many of them. Terraform has JSON-like configuration
files. MongoDB configuration files are in YAML. YCSB\cite{YCSB} configuration
is a Java properties file. We wanted all configuration to be in one place and
in a homogeneous format so that a user could easily understand the whole
configuration.

We also wanted to avoid redundancy in configuration. When Terraform needs
an SSH key to connect to the hosts, and when we later use SFTP to collect
log files, both of these operations should use the same SSH key and it
should exist in the configuration only once.

We wanted maximum flexibility. The framework needed to support any
infrastructure, any possible MongoDB configuration option, any cluster
topology, including deploying more than a single cluster, and needed to
be capable of executing any benchmark software.

The configuration also needed to be modular. It must be possible to execute
the same YCSB test against different MongoDB configurations, and it must be
possible to use the same MongoDB configuration on different hardware.

We also needed to be able to document and trace every bit of configuration.
A common organizational setup is a separate team that develops and deploys
the operating system images. If that team changes the image without the
developers or performance experts noticing, and such changes lead to
performance changes, then much time can be wasted as engineers futilely
try to diagnose a change they do not have visibility into. We use
vanilla OS images available in EC2. All operating system configuration is
done by scripts that are part of DSI. This means that the entire
configuration, and changes to it, can be reviewed from a single commit history.

While developing DSI, we learned that our system performance testing also had
issues with repeatability of test results. So as a parallel project we learned
to configure EC2, Linux and the tests themselves to minimize system noise.
We have reported on this work previously in~\cite{Ingo2019}. DSI
encapsulates the results of that work in its configuration. When MongoDB
engineers use DSI for performance testing, they automatically deploy
systems configured for minimal noise and maximum repeatability.

\subsection{Cloud-first}

In 2016 MongoDB was already heavily invested in using public cloud
infrastructure for testing and CI. We had even developed our own CI system,
Evergreen\cite{Erf2016}, to replace Jenkins\cite{smart2011jenkins} with
a system where everything was built from the ground up to use elastic cloud
resources. As
commits tend to happen during daytime, the need for CI builds varies over
the day. To minimize the turnaround time from commit to a completed CI
build, we try to parallelize test execution as much as possible.

For system performance testing, the elasticity of cloud infrastructure is
even more valuable. We want to test realistic clusters. A weekly build
over a cluster with 3 shards has 16 servers. To test scaling to the largest
instance sizes we occasionally have tested servers with 96 CPUs or 500
GB of RAM. To procure such servers as on-premise hardware would not be
realistic.

At the start of the project, an open question was whether public cloud
infrastructure could be relied on for performance testing at all. We found
when comparing on-premise and cloud that there was no difference in
terms of repeatable performance~\cite{Ingo2019}.

\subsection{Diagnosability}
The primary use case is to run daily CI builds on public cloud servers.
This requires full end-to-end automation.

A challenge in diagnosing regressions and test errors, is that by
the time a human is looking at the results, the servers are no longer
available. For example, we once had a failure where the storage engine
aborted during a test. A possible reason for the abort could simply be
that the disk was full. But we had no way to know whether the disk had
been full.

At the end of the test, DSI therefore collects all the log files
back to the control host, before the test cluster is
terminated. The results are then uploaded to S3. Around 2017 the
MongoDB server had itself
developed rather sophisticated metrics collection. Full Time Diagnostic
Data Capture~\cite{FTDC} collects over a thousand metrics, including system level
metrics like CPU, disk, network, and memory utilization. All we needed
to do was to save those files. To augment these metrics, we also borrowed
 a script used by our support team~\cite{mdiag} which collects additional
information from the host, such as how full the disks are.

An important question, when diagnosing test results, is to verify what
configuration exactly was used for the test. Therefore the DSI input
configuration is added to the same archive file as the logs and test
results. This configuration can also be used to reproduce the regression
if required.

\subsection{Support manual benchmarking outside of CI}
In addition to running benchmarks in Evergreen, DSI also needs to be
usable on its own. This is important both for engineers developing DSI
itself, and for MongoDB engineers who need to reproduce and diagnose performance
issues assigned to them. This requirement may seem obvious, but at the
start of the project, the shell script based system was so entangled with the
evergreen configuration file, that it had become practically impossible
to reproduce tests outside of Evergreen. It was faster to submit a new
CI job than to try to execute the same manually!

As these system-level tests are complex and have many parameters,
commands that did support command line options could often stretch
up to 3 lines. We felt this was poor usability, and in fact went against
the goal of easy reproducibility of tests, as it was easy to miss some
option. As a result, we decided to ban command line options completely, and
force all options into configuration files. This also ensured a full
audit trail, since configuration files would be archived after each
test, while command line options may not be recorded anywhere.

The Evergreen configuration for MongoDB system performance tests is
stored in the MongoDB source code repository~\cite{system_perf_yml},
while DSI itself is its own repository. This often led to situations
where a change had to be committed part to DSI, part to MongoDB, and
in the latter case often to 3 different stable branches too. This
added both unnecessary complexity and stress. A high-priority goal
was for the interface between the two repositories to be as minimal
and stable as possible.

\section{High Level Architecture}
Based on experience from the existing system, we could identify eight
separate modules, that together constitute a full execution of one or
more benchmarks.

\setcounter{subsection}{1}
\subsubsection*{Bootstrap}
As the system began to take shape, we realized that automating some
repetitive setup tasks from the beginning made sense: Copying the
configuration files into place, finding EC2 credentials, installing
correct version of Terraform. This evolved into the bootstrap module.

The \lstinline{bootstrap.yml} configuration file acts as the interface
between the MongoDB repository and the DSI repository. It declares the
composition of configuration files wanted for all the other modules.
For example: YCSB on a three-node replica set. Those other configuration
files are stored inside DSI repository. Thus, the details of each
configuration is abstracted away from the MongoDB repository.

\subsubsection*{Infrastructure provisioning} 
Infrastructure provisioning uses Terraform to deploy EC2 resources.
The DSI configuration, in YAML format, is significantly simpler than
the Terraform files used behind the scenes. It also ensures that if
the deployment fails, the Terraform \textit{destroy} command is called
to cleanup any half-deployed resources.

\subsubsection*{System setup}
As of this writing, system setup was never implemented as its own top
level executable. It remains a "remote script" executed by Terraform.
One drawback is that it is therefore unaware of the rich DSI configuration
files and essentially installs the same software and formats the same
disks each time.

\subsubsection*{Workload setup}
This module installs dependencies for the specific test, such as
Java for YCSB\cite{YCSB} and Linkbench\cite{Linkbench}. An interesting
question was whether workload setup should happen before or after
MongoDB setup. Some tests need to provision a data directory, and this
must be done before MongoDB starts. Other tests need to specify a shard
key, and this must be done against a running MongoDB cluster. In the
end we came to the conclusion that workload\_setup is run first, and
the test control module offers \lstinline{pre_task} and
\lstinline{pre_test} hooks to cover for the second use case.

\subsubsection*{MongoDB setup}
MongoDB setup deploys one or more MongoDB clusters.

\subsubsection*{Test control}
The test control module supports several in-house and third-party
benchmark tools. In addition to running the benchmark itself, it supports
the aforementioned  \lstinline{pre_task} and
\lstinline{pre_test} hooks. After the test it collects a number of log
files and diagnostics back to the control host.

\subsubsection*{Analysis}
Analysis of results comes in two parts. Our main interest is to
determine whether benchmark results deviate from some recent history
of results, and secondarily from past releases. We eventually concluded
that this question is better answered by signal processing algorithms
that look at our daily results holistically as a timeseries, rather than
focusing on the single point in time that is the currently finished
benchmark result. We have reported on this work in~\cite{Daly2020}.
These algorithms are developed and executed separately from DSI.

Some static checks remain in DSI: Does the MongoDB server log
file contain errors
or stack traces, or are there core files on the cluster hosts, etc? The
scope of DSI is therefore everything that concerns a single system
performance task execution. Anything that requires historical data as
input is in the signal processing project.

\subsubsection*{Infrastructure teardown}
Infrastructure teardown uses Terraform to release the cloud resources.

The modules should be executed in the above order. To allow for flexibility
and modularity, it is however allowed to skip some. For example, a user
who wants to test an existing MongoDB cluster could just point
test\_control directly at that.

\section{Implementation}
The analysis code was already written in Python. We ended up writing the
other modules in Python as well. Maybe it was mostly as a function of
gravity, but we have found Python a rather suitable language for this
task. The combination of being a real programming language and yet a
scripting language that does not require compilation has afforded a lot
of flexibility during the active development phase.

\subsection{Configuration}
The configuration files are the heart of DSI. Since we do not allow
command line options at all, the configuration files are essentially
the user interface. The \lstinline{docs/} directory in the source code~\cite{DSIRepo}
documents all the available configuration options. The
\lstinline{configurations/} directory contains a growing selection of
canned configurations, including all configurations used by daily CI builds.

As mentioned in the design goals (see \ref{sec:problem-configuration}),
ideally we wanted all configuration in one big file, but modularity
requires that users can mix and match different configurations. The
actual implementation has one YAML configuration file for each module: infrastructure\_provisioning.yml, workload\_setup.yml, etc...

To ensure that all modules read the configuration in a consistent way,
a library is used to access the configuration. The library reads all
the YAML files into one big Python dictionary structure. So from the
DSI developer's perspective it truly is like all configuration was
provided from a single big YAML file. The same library also provides
various additional functionality. All defaults are centralized in a
\lstinline{defaults.yml} file, from which a default value is transparently
returned when the actual config file does not specify a value. Since YAML
anchors cannot be used between different YAML files, the library also
provides a \$\{variable.reference\} syntax to reference one configuration
value from another part of the configuration.

\begin{lstlisting}[caption={mongodb\_setup.yml}, numbers=left, basicstyle={\tiny\ttfamily}, label=lst:mongodb, float=tb]
mongod_config_file:
  storage:
    engine: wiredTiger
  replication:
    replSetName: rs0

topology:
  - cluster_type: replset
    id: rs0
    mongod:
      - public_ip: ${infrastructure_provisioning.out.mongod.0.public_ip}
      - public_ip: ${infrastructure_provisioning.out.mongod.1.public_ip}
      - public_ip: ${infrastructure_provisioning.out.mongod.2.public_ip}

# Meta data about this mongodb setup
meta:
  # The list of hosts that can be used in a mongodb connection string
  hosts: ${mongodb_setup.topology.0.mongod.0.private_ip}:27017
  hostname: ${mongodb_setup.topology.0.mongod.0.private_ip}
  mongodb_url: mongodb://${mongodb_setup.meta.hosts}/test?replicaSet=rs0
  is_replset: true
\end{lstlisting}

\begin{lstlisting}[caption=test\_control.yml, numbers=left, basicstyle=\tiny\ttfamily, label=lst:test, float=tb]
run:
  - id: ycsb_load
    type: ycsb
    cmd: ./bin/ycsb load mongodb -s -P ../../workloadEvergreen -threads 8
    config_filename: workloadEvergreen
    workload_config: |
      mongodb.url=${mongodb_setup.meta.mongodb_url}
      recordcount=5000000
      workload=com.yahoo.ycsb.workloads.CoreWorkload
  - id: ycsb_100read
    type: ycsb
    cmd: ./bin/ycsb run mongodb -s -P ../../workloadEvergreen_100read -threads 32
    config_filename: workloadEvergreen_100read
    workload_config: |
      mongodb.url=${mongodb_setup.meta.mongodb_url}
      recordcount=5000000
      maxexecutiontime=240
      workload=com.yahoo.ycsb.workloads.CoreWorkload
      readproportion=1.0
\end{lstlisting}

Listings \ref{lst:mongodb} and \ref{lst:test} show two simplified
configuration files. Keys \lstinline{mongod_config_file} and \lstinline{workload_config}
embed literal MongoDB and YCSB configuration files. When executing
the benchmark, DSI will extract these into their own files, that are
used as input to \lstinline{mongod} and \lstinline{ycsb} respectively.

Since we do not know the IP address of an EC2 host ahead of time, the
infrastructure\_provisioning module will store that information in a
special \textit{.out.yml} file. The MongoDB topology configuration uses
those IP addresses via references. The config library
resolves these automatically.

The \lstinline{meta} section facilitates the modular mixing of config
files. Since the MongoDB URL depends on the type of MongoDB cluster, the
URL to use is provided together with the MongoDB configuration. The YCSB
configuration then uses the provided URL via reference.

The system also supports a variety of hooks for setup and post processing.
For example, a \lstinline{pre_cluster_start} hook can be used to download
and provision database files before starting MongoDB, and a
\lstinline{pre_task} hook might create indexes or shard collections before
the test starts.

For a consistent experience, all configuration is in YAML. Terraform
and MongoDB Atlas\footnote{Atlas is MongoDB's Database as a Service product}
API expect JSON syntax as input. DSI transparently
transforms the YAML configuration for those components to JSON.

\subsection{Technical Challenges}
We have been early adopters of Terraform~\cite{Terraform}. Generally it
has served us well, but some details have required workarounds along the
way. Early on Terraform had a hard limit for how many servers it could
provision at once. As a workaround we needed to deploy our sharded cluster
in two separate steps. Another problem was that the plugin that configures
AWS Virtual Private Cloud (VPC) networks lacked dependency information.
This caused server deployments to fail when they tried to use a VPC that
did not exist yet.

Newer Terraform versions almost never supported the configuration files
from the older versions. This caused us to upgrade only rarely, as it was
often days of work.

Originally we reused clusters between Evergreen tasks. This
was implemented because EC2 billing was hourly, and we wanted to avoid
situations where we would have to double pay for an hour due to terminating and
instantly re-deploying the same cluster. As EC2 moved to per minute
billing, we wanted to remove this code path that is used in CI only
and therefore hard to test and easy to forget. Yet doing this caused
\lstinline{InsufficientInstanceCapacity} errors from EC2. It may have
been related to some
special EC2 configurations we use: either dedicated instances or placement
groups. For whatever reason it was not possible to terminate a
cluster of N nodes and then expect to immediately start the same N
nodes again. Recently we revisited this issue, and it seems this is no
longer an issue. So now we were finally able to remove the re-use of clusters.

\subsection{Human Challenges}
Many challenges are human rather than technical:

DSI takes a \textit{laissez faire} attitude to all the configuration.
DSI's role is pass thru in nature: YAML is turned into a Python dictionary structure,
part of which is passed as input to \lstinline{mongod} and \lstinline{ycsb}.
If there is an error in the configuration, then at this point
\lstinline{mongod} or \lstinline{ycsb} will print an error and fail.
Trying to duplicate such an error checking procedure in DSI would be
futile: MongoDB supports hundreds of configuration options and new
ones are added every year.

Yet software engineers are trained to validate user input. Teaching a
software team to not do so is surprisingly hard. Early on a module had
been written in Go, a strongly typed language. Each time we wanted to
pass a new option to the test, we had to add an attribute to the Go class
that held the JSON input. This was a huge impediment until the Go code
was replaced by Python.

A similar struggle has been around the issue of providing defaults
from a centralized \lstinline{defaults.yml} file. The instinct to
always provide a default \textit{in the code} --- such as by sprinkling
Python constants around the Python code --- is strong among well educated
software engineers. Accepting that there is a default, it's just
provided by a library and therefore not visible in the lines of code
being written at this moment, can be hard. Yet in the DSI architecture,
providing default values in the code can be considered an error. Since
different modules access the same configuration, it is possible (and
encouraged) that two independent code paths can access the same
configuration option. If both of them specify default values in
the Python code, there is a chance that they specify different defaults,
which would be inconsistent and a bug.

A final observation around challenges of the human mind has been a
recurring tendency to overestimate the importance of whatever small
feature an engineer is currently working on. This recency bias causes engineers
to add new options to \lstinline{bootstrap.yml} --- the top level
configuration file end users look at first --- and documenting this option
on the first page of the user manual. But DSI has hundreds of
configuration options and putting all of them in the top level is not
scalable. Moving options from the \lstinline{bootstrap.yml} config file
further "down" into other, more topical sections of the configuration,
remains a regular occupation for the authors.

\section{Results}
\begin{figure}[h]
  \centering
  \includegraphics[width=\linewidth]{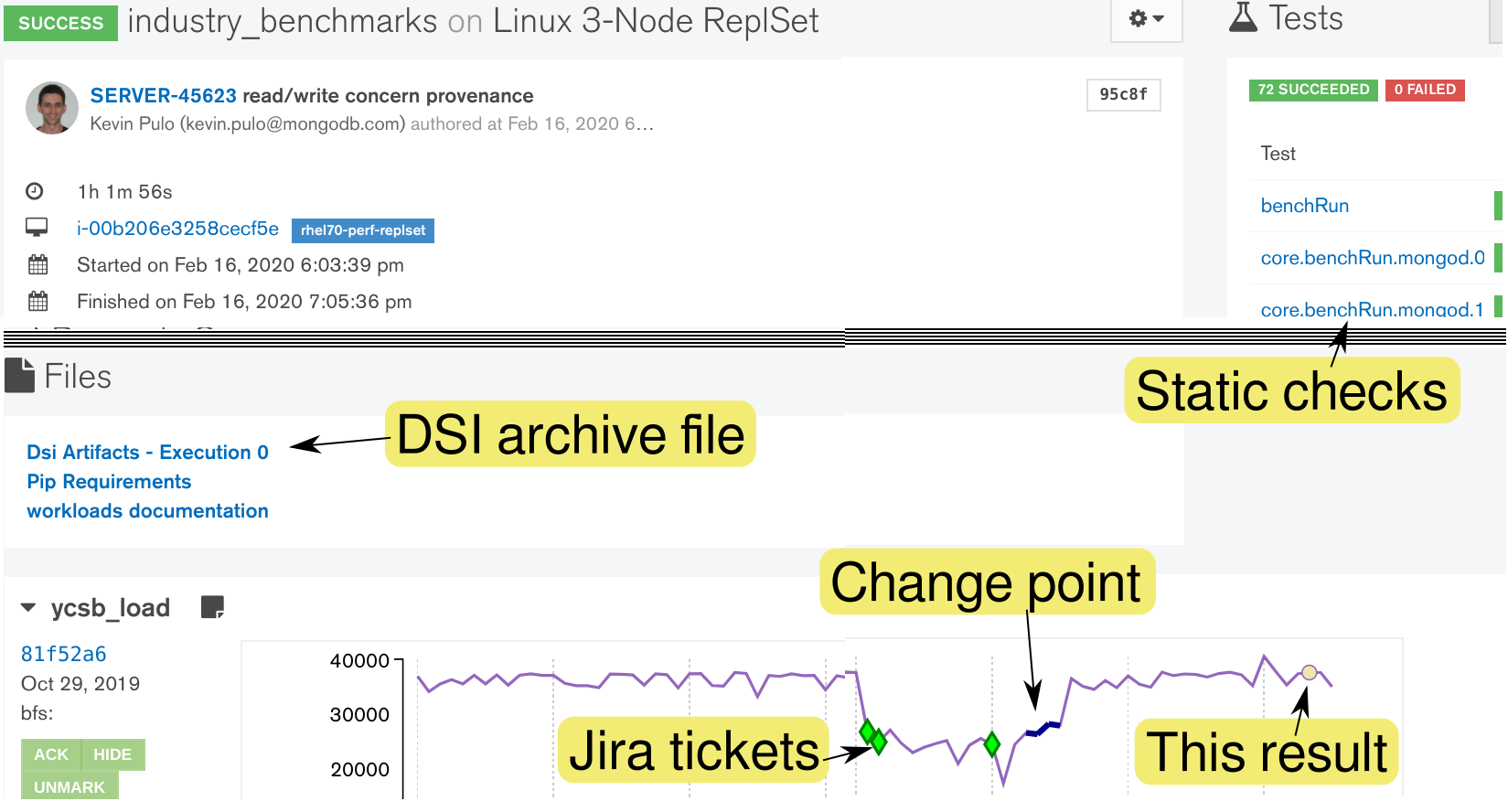}
  \caption{Results for a YCSB test in the Evergreen CI environment.}
  \Description{UI from a performance test result in Evergreen CI.}
  \label{fig:evergreen}
\end{figure}

Figure \ref{fig:evergreen} shows the Evergreen result page of a task
where DSI has executed the YCSB benchmark. The benchmark results are
shown as a yellow dot on the graphs in the bottom. The graph as a whole
is the timeseries of daily build results. Bolded lines are statistically
significant changes highlighted by the signal processing algorithm, and
the green diamonds link to associated Jira tickets. The investment
in running performance tests daily in CI has paid off: A major regression
has been found and fixed as part of normal engineering process, instead
of waiting until a release candidate is released and only finding issues
then. The top right corner shows fail/pass status of the static checks
done by the DSI analysis module.

As the development of DSI proceeded and its usability improved, the
system performance benchmarking project in Evergreen has become the
primary target for MongoDB engineers to write performance tests for
new features. As of March 2020 we have almost 200 tests and 20 MongoDB
configurations in the DSI repository, most of which run once per day. This
is on top of the single-node microbenchmarks that have been in use before
DSI, and also in addition to the Google Benchmark C++ unit tests, which
run several times per day.

The goal of developing a flexible system that could be used to test
arbitrary MongoDB configurations was achieved. An early validation
of the design was the addition of a test to measure performance of the
initial synchronization of data files, when a new node is added to a
replica set. The test
required one of the MongoDB nodes to be deployed detached from
the replica set and in uninitialized state. This abnormal setup was
possible with a configuration change only. Similar small victories,
such as using arbiter nodes in a replica set, have followed later.

While we have focused our testing on a single platform: EC2 and Amazon
Linux, the goal was for DSI to be able to support also other hardware
infrastructure. This was finally proven recently when we have pointed
a test\_control.yml file against servers running in Azure. A similar
triumph of modular design was to replace the deployment of a MongoDB cluster
with a HTTP call to deploy an Atlas
cluster instead. By replacing mongodb\_setup entirely, DSI could be used
to test other database products than MongoDB. Similarly, to add more
infrastructure options, such as Kubernetes, we would add support
for them in infrastructure\_provisioning module.

Reproducing regressions caught by the CI system is straightforward. An
engineer can simply download the "DSI Artifacts" tar file from the Evergreen
result page (see \ref{fig:evergreen}), untar the package, cd into the
directory, and execute the sequence of DSI commands
(\lstinline{infrastructure_provisioning}, ...) to reproduce the exact
same steps as were executed in CI.

During the MongoDB 4.2 development cycle the system performance 
CI builds caught 63 regressions,
far beyond the microbenchmarks project at 20. But more importantly, the
MongoDB Server engineers are now committed and able to reproduce and
investigate the regressions from CI. As a result, all but one of the
major performance regressions had been fixed before stable release.
In addition, we are also able to track net new
improvements in performance, of which there were 17 caught by DSI tests.
The major improvements were chronicled by marketing in a blog
post~\cite{DJ2019}.



\section{Related Work}
At the time we started developing DSI, we were not aware of any similar
end-to-end tools for performance testing of distributed databases --- and
certainly not one supporting MongoDB.

On first sight, it still seems like not much has been published on the
topic of automating system level performance benchmarking. An
illustrative example is the CockroachDB documentation, which includes
a guide for benchmarking CockroachDB with TPC-C~\cite{CockroachTPCC}.
It directs the user to start with manually deploying a CockroachDB
cluster: \textit{"Repeat steps 1 -- 3 for the other 80 VMs for CockroachDB
nodes".} In reality this is not how Cockroach Labs engineers test
CockroachDB clusters, rather the Github repository does include a
DSI-like pair of tools \textit{roachprod} and
\textit{roachtest}~\cite{roachtest}.

Mikael Ronstr\"{o}m's book "MySQL Cluster 7.5 inside and out" includes
a whole chapter on \lstinline{bench_run.sh}, which supports several
benchmark clients used to benchmark MySQL NDB
Cluster~\cite{ronstrom2018mysql}. It requires the user to have downloaded
the necessary software to the control host, and expects the servers to
exist, but from this starting point automates deployment of the MySQL
software and execution of the test.

Scylla Cluster Tests (SCT)~\cite{ScyllaClusterTests} is an end to
end framework very similar to DSI. It supports the use of a "test
oracle", a reference database that executes the same tests and is
assumed to always produce the correct response. The test oracle
could be a previous stable release.

Rally is the benchmarking tool for ElasticSearch~\cite{ElasticRally}.
It is installable as a Python module via the \textit{pip} tool,
and comes with an extensive user manual.
Similar to SCT, it supports benchmarking different ElasticSearch releases
and comparing the results. Rally does not deploy any cloud resources,
rather expects cluster hosts to already exist.

The above approach to compare a new version in parallel to a "known
good" release seems to be a common approach. While DSI does not support
that as a first class feature, we should note that we do similar
comparisons of the development branch results against recent stable
releases. However, we consider these monthly and annual reviews as a
stopgap measure only. We feel that in a daily CI oriented workflow,
focusing on comparing results against a recent history helps us catch
regressions as they happen.

SAP HANA uses a more CI oriented workflow for performance testing, very
similar to ours~\cite{Rehmann2016}. At MongoDB we have recently moved
to a process, where merges to the master source code branch go through an
automated CI gatekeeper. However, we have yet to add performance
tests to this automated gatekeeping process, rather engineers must submit
their changes for performance testing on an opt-in basis. SAP is using
performance tests as part of their gatekeeping step. SAP reports also
resource consumption, such as CPU utilization, as part of their test
results. They note that only a human evaluator can judge
whether increased CPU utilization is a positive or negative change.

Outside of the database space, BenchFlow~\cite{Ferme2017}
provides a very DSI-like end-to-end automation for benchmarking web
services.

\section{Conclusions}
Distributed Systems Infrastructure (DSI) is MongoDB's framework for
running fully automated system performance tests in our CI environment.
Automating deployment of real multi-node clusters, executing tests,
tuning the system for repeatable results, and then collecting and
analyzing the results, is a hard problem, and it took us 3 attempts
and 6 years to get it right. The open sourced DSI project is the
result of those efforts.

Today DSI is MongoDB's most used and most useful performance testing tool.
It runs almost 200 different benchmarks in daily CI, and we also use it
for manual performance investigations. During MongoDB 4.2 development cycle,
DSI caught 63 regressions. As we can alert the responsible engineer in
a timely fashion, all but one non-trivial regressions were fixed before
the 4.2.0 release. We are also able to detect net new improvements, of
which DSI caught 17.

DSI presents some novel design choices. For example, we have banned the
use of command line options completely. By forcing all configuration
into configuration files, we get an audit trail of each test execution,
and engineers are also easily able to reproduce the exact same execution.
Another uncommon choice is that we use vanilla images when deploying
servers instead of building our own custom image. This means that all
configuration code is in the DSI repository.

\begin{acks}
Chung-Yen Chang and Rui Zhang were with us in the original Performance
team that started work on a properly designed
Python based version. Jim O'Leary, Ryan
Timmons, and Julian Edwards were part of the team that wrote most of
the code that is in DSI today. Several interns and new grads contributed
major components while visiting the team: Shane Harvey created
\lstinline{mongodb_setup.py} and the SSH plumbing. Ryan Chipman created
\lstinline{bootstrap.py} and Blake Oler the key feature to restart MongoDB
between tests. Severyn Kozak, William Brown, Kathy Chen, Pia Kochar worked
on rules in analysis.py and test configurations. After the active
development phase completed, maintainership of DSI has been with Max
Hirschorn, Ryan Timmons, Robert Guo and Raiden Worley, while we have
taken the role of end users using DSI to investigate MongoDB performance.
It was Ryan Timmons who never gave up the hope of one day open sourcing DSI,
and played an active role in facilitating merges of remaining cleanup work
to make that hope a reality.

Dan Pasette and Ian Whalen have acted as project leads during transition
phases between above team compositions, and overall deserve thanks for not
giving up on the project during a long but necessary rewrite. Cristopher
Stauffer and April Schoffer have provided similar project oversight in
recent years.

We thank Eoin Brazil and the anonymous reviewers for valuable feedback
on this article, helping us to present our work more clearly.
\end{acks}

\bibliographystyle{ACM-Reference-Format}
\bibliography{bibfile}

\end{document}